\documentclass[a4paper]{spie}  
\usepackage{graphicx}
\usepackage{subcaption}
\usepackage{multirow}
\usepackage{amsmath,amsfonts,amssymb}
\usepackage{graphicx}
\usepackage[colorlinks=true, allcolors=blue]{hyperref}

\title{WALOP-South: A wide-field one-shot linear optical polarimeter for PASIPHAE survey}

\author[a*]{Siddharth Maharana}
\author[b,c]{John A. Kypriotakis}
\author[a,b,e]{A. N. Ramaprakash}
\author[a]{Pravin Khodade}
\author[a]{Chaitanya Rajarshi}
\author[a]{Bhushan Joshi}
\author[a]{Pravin Chordia}
\author[a]{Ramya M. Anche}
\author[h]{Shrish}
\author[b,c,i]{Dmitry Blinov}
\author[g]{Hans Kristian Eriksen}
\author[h]{Tuhin Ghosh}
\author[g]{Eirik Gjerløw}
\author[b,c]{Nikolaos Mandarakas}
\author[f]{Georgia V. Panopoulou}
\author[b,c]{Vasiliki Pavlidou}
\author[e]{Timothy J. Pearson}
\author[b,c]{Vincent Pelgrims}
\author[d,j]{Stephen B. Potter}
\author[e]{Anthony C. S. Readhead}
\author[b,c]{Raphael Skalidis}
\author[b,c]{Konstantinos Tassis}
\author[g]{Ingunn K. Wehus}

\affil[a]{Inter-University Centre for Astronomy and Astrophysics, Post bag 4, Ganeshkhind, Pune, 411007, India}
\affil[b]{Institute of Astrophysics, Foundation for Research and Technology-Hellas, Voutes, 70013 Heraklion, Greece}

\affil[c]{Department of Physics, University of Crete, Voutes, 70013 Heraklion, Greece}

\affil[d]{South African Astronomical Observatory, PO Box 9, Observatory, 7935, Cape Town, South Africa}

\affil[e]{Cahill Center for Astronomy and Astrophysics, California Institute of Technology, Pasadena, CA, 91125, USA}

\affil[f]{Hubble Fellow, California Institute of Technology, Pasadena, CA 91125, USA}

\affil[g]{Institute of Theoretical Astrophysics, University of Oslo, P.O. Box 1029 Blindern, NO-0315 Oslo, Norway}

\affil[h]{School of Physical Sciences, National Institute of Science Education and Research, HBNI, Jatni 752050, Odisha, India}

\affil[i]{Astronomical Institute, St. Petersburg State University, 198504, St. Petersburg, Russia}

\affil[j]{Department of Physics, University of Johannesburg, PO Box 524, Auckland Park 2006, South Africa}

\authorinfo{Further author information: (Send correspondence to S.M.)\\S.M.: E-mail: sidh@iucaa.in, Telephone: +91 020 2560 4213}

\begin{document} 
\maketitle

\begin{abstract}
WALOP (Wide-Area Linear Optical Polarimeter)-South, to be mounted on the 1m SAAO telescope in South Africa, is first of the two WALOP instruments currently under development for carrying out the PASIPHAE survey. Scheduled for commissioning in the year 2021, the WALOP instruments will be used to measure the linear polarization of around $10^{6}$ stars in the SDSS-r broadband with 0.1~\% polarimetric accuracy, covering 4000 square degrees in the Galactic polar regions. The combined capabilities of one-shot linear polarimetry, high polarimetric accuracy ($< 0.1~\%$) and polarimetric sensitivity ($< 0.05~\%$), and a large field of view (FOV) of $35\times35$~arcminutes make WALOP-South a unique astronomical instrument. In a single exposure, it is designed to measure the Stokes parameters $I$, $q$ and $u$ in the SDSS-r broadband and narrowband filters between 500-700~nm. During each measurement, four images of the full field corresponding to the polarization angles of $0^{\circ}$, $45^{\circ}$, $90^{\circ}$ and $135^{\circ}$ will be imaged on four detectors and carrying out differential photometry on these images will yield the Stokes parameters. Major challenges in designing WALOP-South instrument include- (a) in the optical design, correcting for the spectral dispersion introduced by large split angle Wollaston Prisms used as polarization analyzers as well as aberrations from the wide field, and (b) making an optomechanical design adherent to the tolerances required to obtain good imaging and polarimetric performance under all temperature conditions as well as telescope pointing positions. We present the optical and optomechanical design for WALOP-South which overcomes these challenges.
\end{abstract}


\section{INTRODUCTION}\label{sec:intro}
WALOPs (Wide-Area Linear Optical Polarimeter) are a pair of wide field linear optical polarimeters currently under development at the \href{https://instru.iucaa.in/}{Inter-University Center for Astronomy and Astrophysics (IUCAA)}, Pune, India. WALOP-South will be installed at the  \href{https://www.saao.ac.za/astronomers/1-0m/}{1~m telescope} of South African Astronomical Observatory's Sutherland Observatory while WALOP-North will be installed on the \href{http://skinakas.physics.uoc.gr/en/telescopes/tel_130.html}{1.3~m telescope} of the Skinakas Observatory of the University of Crete, Greece. Together, these will be deployed to carry out the \href{http://pasiphae.science/}{PASIPHAE survey}\cite{tassis2018pasiphae}, which aims to cover 4000 square degrees of sky in the northern and southern Galactic polar regions and measure polarization of around $10^{6}$ stars with polarimetric accuracy better than 0.1~\%. Current optical polarization catalogues have measurement of $10^{4}$ stars\cite{Heiles_2000}. The main scientific objective of the PASIPHAE survey is to use this high accuracy stellar polarimetric survey data in conjugation with the GAIA survey's stellar distance measurements to create a 3-d tomography map of the dust and magnetic field in Milky Way Galaxy's polar regions\cite{Panopoulou_2019}. A detailed description of the scientific motivations and objectives of the PASIPHAE survey is presented in PASIPHAE program's white paper\cite{tassis2018pasiphae}. Of the two WALOP instruments, WALOP-South is scheduled first for commissioning in the year 2021. In this manuscript, we present the optical and optomechanical design of the instrument as well as its current status. Section~\ref{tech_goals} describes the overall design goals of the instrument, while Sections~\ref{optical_design} and \ref{optomech} describe the optical and optomechanical design of the WALOP-South instrument.

\subsection{Design Goals of WALOP-South}\label{tech_goals}

The unique science goals of the PASIPHAE survey drive the technical design goals for the WALOP instruments, which are same for both WALOP-North and WALOP-South, and are listed in Table~\ref{techtable}. The motivation and justification for the target values for each of the design parameters is provided in the dedicated optical design paper of the instrument by Maharana et al. (accepted for publication in Journal for Astronomical Telescopes, Instruments and Systems), henceforth referred to as Paper I.

We define polarimetric sensitivity ($s$) as the least value and change of linear polarization which the instrument can measure, without correction for the cross-talk and instrumental polarization of the instrument. $s$ is a measure of the internal noise and random systematic of the instrument due to the optics. Polarimetric accuracy ($a$) is the measure of closeness of the predicted polarization of a source to the real value after applying corrections for the above effects using calibration techniques (described in Paper I) as well as taking into account uncertainty due to photon noise.

\begin{table}[htbp!]
    \centering
    \begin{tabular}{|c|c|c|}
        \hline
        \textbf{Sl. No}. & \textbf{Parameter} & \textbf{Technical Goal} \\
         \hline
        1 & Polarimetric Sensitivity & 0.05~\%\\
         \hline
        2 & Polarimetric Accuracy & 0.1~\%\\
         \hline
        3 & Polarimeter Type & Four Channel One-Shot Linear Polarimetry\\
         \hline
        4 & Number of Cameras & 4 (One Camera for Each Arm)\\
         \hline
        5 & Field of View & $30\times30$~arcminutes\\
         \hline
        6 & Detector Size & $4k\times4k$ (Pixel Size = $15~{\mu}m$) \\
        \hline
        7 & No. of Detectors & 4 \\
        \hline
        8 & Primary Filter & SDSS-r \\
        \hline
        9 & Imaging Performance & Close to seeing limited PSF \\
         \hline
        9 & Stray and Ghost Light Level & Brightness less than sky brightness per pixel.\\
        \hline
    \end{tabular}
    \caption{Design goals for WALOP-South instrument.}
    \label{techtable}
\end{table}

\section{Optical Design}\label{optical_design}
A major challenge in the development of the WALOP-South instrument was creating a suitable optical design which meets the requirements listed in Table~\ref{techtable}. Here we present the overall optical model of the instrument and it's predicted performance. While two/four channel optical polarimeters with imaging of the two/four channels on different detector/detector-areas have been made in the past, either they have been designed for very narrow fields of view\cite{robopol,HOWPol} of around $1\times1$~arcminutes, or they have not been designed to work for broadband filters\cite{salt_commisioning} (width of $ > = 100~nm$). WALOP-South is first of its kind wide-field one-shot four channel imaging polarimeter with separate cameras for each channel. WALOP-South's optical design has been created to work optimally for the 1~m SAAO telescope's optics prescription and Sutherland Observatory's temperature and observing conditions, which are listed in Table~\ref{telescope_details}.

\begin{table}[htbp!]
    \centering
    \begin{tabular}{|c|c|}
    \hline
    \textbf{Parameter} & \textbf{Value} \\
    \hline
    Telescope Type & Cassegrain Focus and Equatorial Mount \\
    \hline
    Primary Mirror Diameter & 1~m \\
    \hline
    Secondary Mirror Diameter & 0.33~m \\
    \hline
    Nominal Telescope f-Number & 16.0 \\
    \hline
    Altitude & 1800~m\\
    \hline
    Median Seeing FWHM & 1.5" \\
    \hline
    Extreme Site Temperatures & $-10^{\circ}~C$ to $40^{\circ}~C$\\
    \hline
    \end{tabular}
    \caption{Telescope and site details of South African Astronomical Observatory's (SAAO) Sutherland Observatory.}
    \label{telescope_details}
\end{table}

\begin{figure}
    \centering
    \fbox{\includegraphics[scale=0.45]{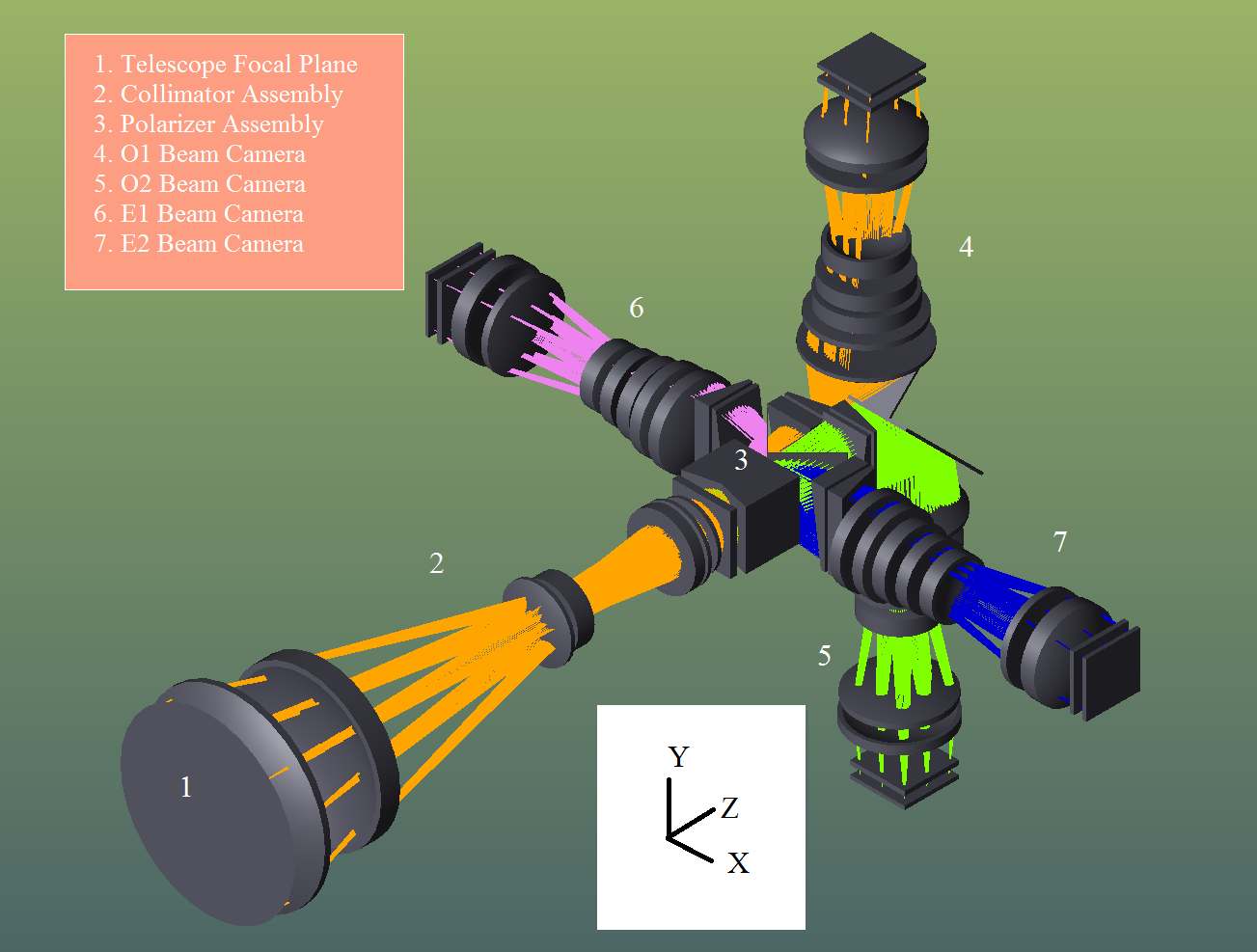}}
    \caption{Optical model of the WALOP-South instrument. Beginning at the telescope focal plane, it accepts the beam for the entire field of view and through the collimator assembly creates a pupil image, which is then fed to the polarizer assembly. The polarizer assembly acts as the polarization analyzer system of the instrument and splits the pupil beam into four channels corresponding to $0^{\circ}$, $45^{\circ}$, $90^{\circ}$ and $135^{\circ}$ polarization angles, which are referred to as O1, O2, E1 and E2 beams, respectively. Additionally, this assembly folds and steers the O beams along the +y and -y directions and the E beams along the +x and -x directions. Each channel has its own camera to image the entire field of view on a $4k\times4k$ CCD detector.}
    \label{WALOP-S}
\end{figure}

The optical model of WALOP-South was designed and analyzed using the \href{https://www.zemax.com/}{Zemax}\textsuperscript{\textregistered} optical design software. Figure~\ref{WALOP-S} shows the optical model of the instrument. The entire instrument's optical system consists of the following assemblies: a collimator, a polarizer assembly and four cameras (one for each channel). The collimator assembly begins from the telescope focal plane. Aligned along the z-axis, it creates a pupil image which is fed to the polarizer assembly. The polarizer assembly acts as the polarization analyzer system of the instrument and splits the pupil beam into four channels corresponding to $0^{\circ}$, $45^{\circ}$, $90^{\circ}$ and $135^{\circ}$ polarization angles, which are referred to as O1, O2, E1 and E2 beams, respectively. Additionally, this assembly folds and steers the O beams along the +y and -y directions and the E beams along the +x and -x directions. Each channel has its own camera to image the entire field of view on a $4k\times4k$ CCD detector. The obtained field of view of the instrument is $34.8\times34.8$~arcminutes, although the required field of view was $30\times30$~arcminutes. Table~\ref{op_design_summary} lists the key design parameters of the instrument's optical system. The polarizer assembly the most novel and complex aspect of the WALOP-South optical design, and Section~\ref{pol_ass} describes it's architecture and working. As part of the optical design, we also designed a guider camera for instrument as well as new baffles for the telescope to accommodate the large field of view of WALOP-South- these are described in Paper I.

\begin{table}[htbp!]
    \centering
    \begin{tabular}{|c|c|}
        \hline
        \textbf{Parameter} & \textbf{Design Value/Choice}\\
        \hline
        Filter & SDSS-r \\
        \hline
        Telescope F-number & 16.0 \\
        \hline
        Camera F-number & 6.1 \\
        \hline
        Collimator Length & 700~mm\\
        \hline
        Camera Length & 340~mm \\
        \hline
        No of lenses in Collimator & 6 \\
        \hline
        No of lenses in Each Camera & 7 \\
        \hline
        Detector Size & $4096\times4096$ \\
        \hline
        Pixel Size & $15~{\mu}m$ \\
        \hline
        Sky Sampling at detector & 0.5"/pixel\\
        \hline
        
    \end{tabular}
    \caption{Values of the key parameters of WALOP-South Optical Design.}
    \label{op_design_summary}
\end{table}

\subsection{Polarizer Assembly Design}\label{pol_ass} 
\begin{figure}
    \centering
    \frame{\includegraphics[scale = 0.25]{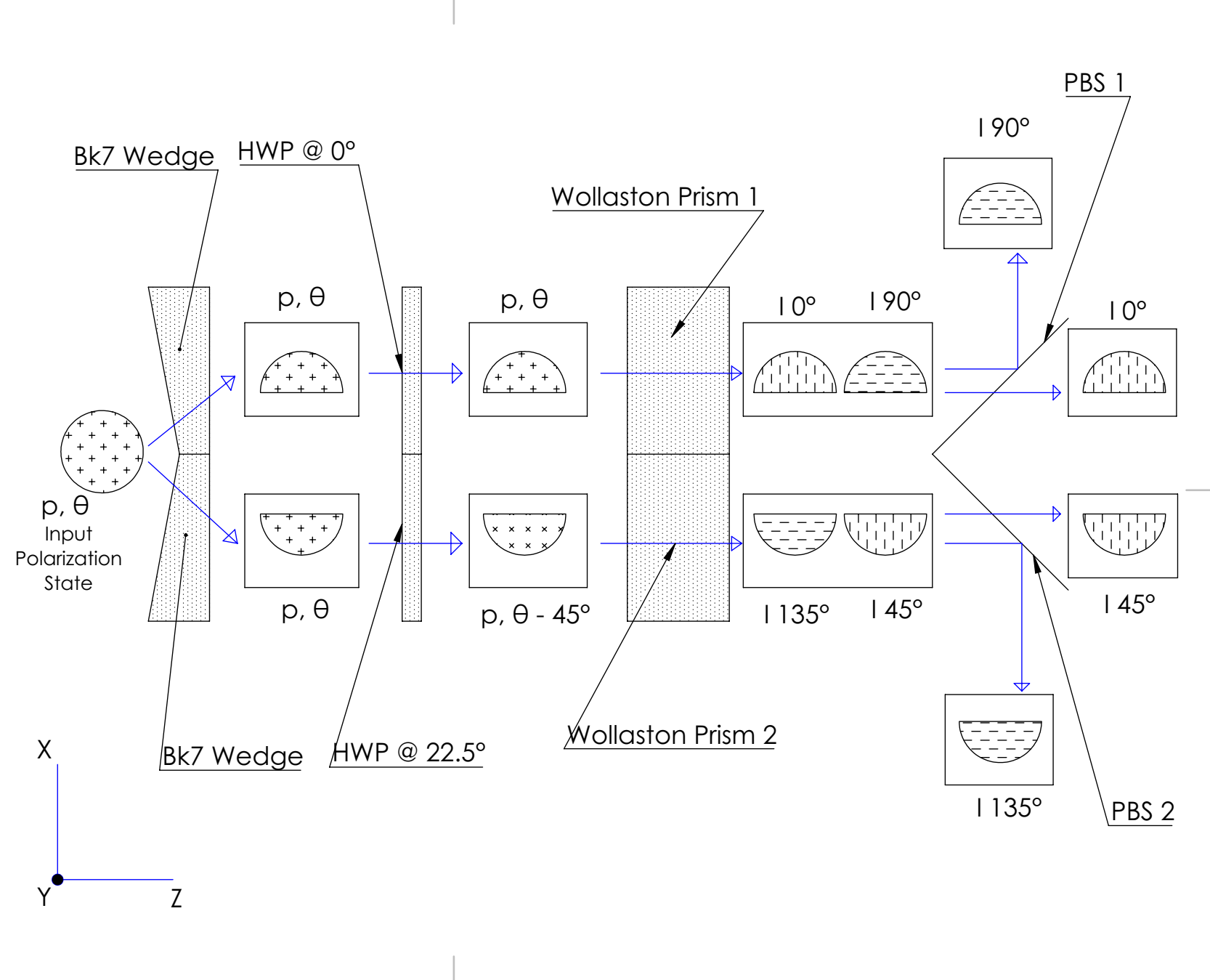}}
    \caption{A cartoon illustrating the working of the polarizer assembly of the WALOP-South instrument. $p$ and $\theta$ shown are as seen in the x-y plane when viewed along the z-axis of the cartoon and the change in the polarization state of the beams while passing through this system is annotated. Together, the Wollaston Prism Assembly consisting of the two BK7 glass wedges, Wollaston Prisms (WP) and Half-Wave Plates (HWP) and the two PBS' act as the polarization beamsplitter sub-system. The pupil is split between the two BK7 wedges which is then fed to the twin HWP + WP system to be split into four channels with the polarization states of $0^{\circ}$, $45^{\circ}$, $90^{\circ}$ and $135^{\circ}$. Afterwards, the two PBS' direct these four beams in four directions. }
    \label{pol_ass_cartoon}
\end{figure}

It consists of four sub-assemblies: (a) Wollaston Prism Assembly (WPA), (b) Wire-Grid Polarization Beam-Splitter (PBS), (c) Dispersion Corrector Prisms (DC Prisms) and (d) Fold Mirrors. 

The WPA consists of two identical calcite Wollaston Prisms (WP), with a half-wave plate (HWP) and a BK7 glass wedge in front of each WP (Figure~\ref{pol_ass_cartoon}). The WPs have an aperture of $45\times80~mm$ and a wedge angle of $30^{\circ}$, resulting in a split angle of $11.4^{\circ}$ at $0.6~{\mu}m$ wavelength. The left WP has a HWP with fast-axis at $0^{\circ}$ with respect to the instrument coordinate system to separate $0^{\circ}$ and $90^{\circ}$ polarizations while the right WP has a HWP with fast-axis at $22.5^{\circ}$ to split the $45^{\circ}$ and $135^{\circ}$ polarizations. The BK7 wedges at the beginning of the WPA, which share the incoming pupil beam equally, ensure that rays from the off-axis objects in the field of view entering at oblique angles of incidence do not hit the interface between the WPs, which will lead to throughput loss as well as instrumental polarization from scattering arising at the surface. Thus the WPA, using the splitting action of the WPs, separates the beam at the pupil into O1, O2, E1 and E2 beams- corresponding to the  polarization angles of $0^{\circ}$, $45^{\circ}$, $90^{\circ}$ and $135^{\circ}$ respectively. The PBS' act as beam selectors, allowing both the O beams to pass through while folding the E1 and E2 beams along -x and +x directions. Figure~\ref{pol_ass_cartoon} shows the overall working idea of the WPA and PBS components of the polarizer assembly. The DC Prisms are a pair of glass prisms present in the path of each of the four beams after the PBS' to correct for the spectral dispersion introduced by the WPA (refer to Paper I). Additionally, mirrors placed at $\pm~45^{\circ}$ the y-z plane fold the O beams into +y and -y directions to limit the length of the instrument to 1.1~m from the telescope focal plane.

\subsection{Optical Performance and Tolerance Analysis}

\begin{figure}
    \centering
    \includegraphics[scale = 0.15]{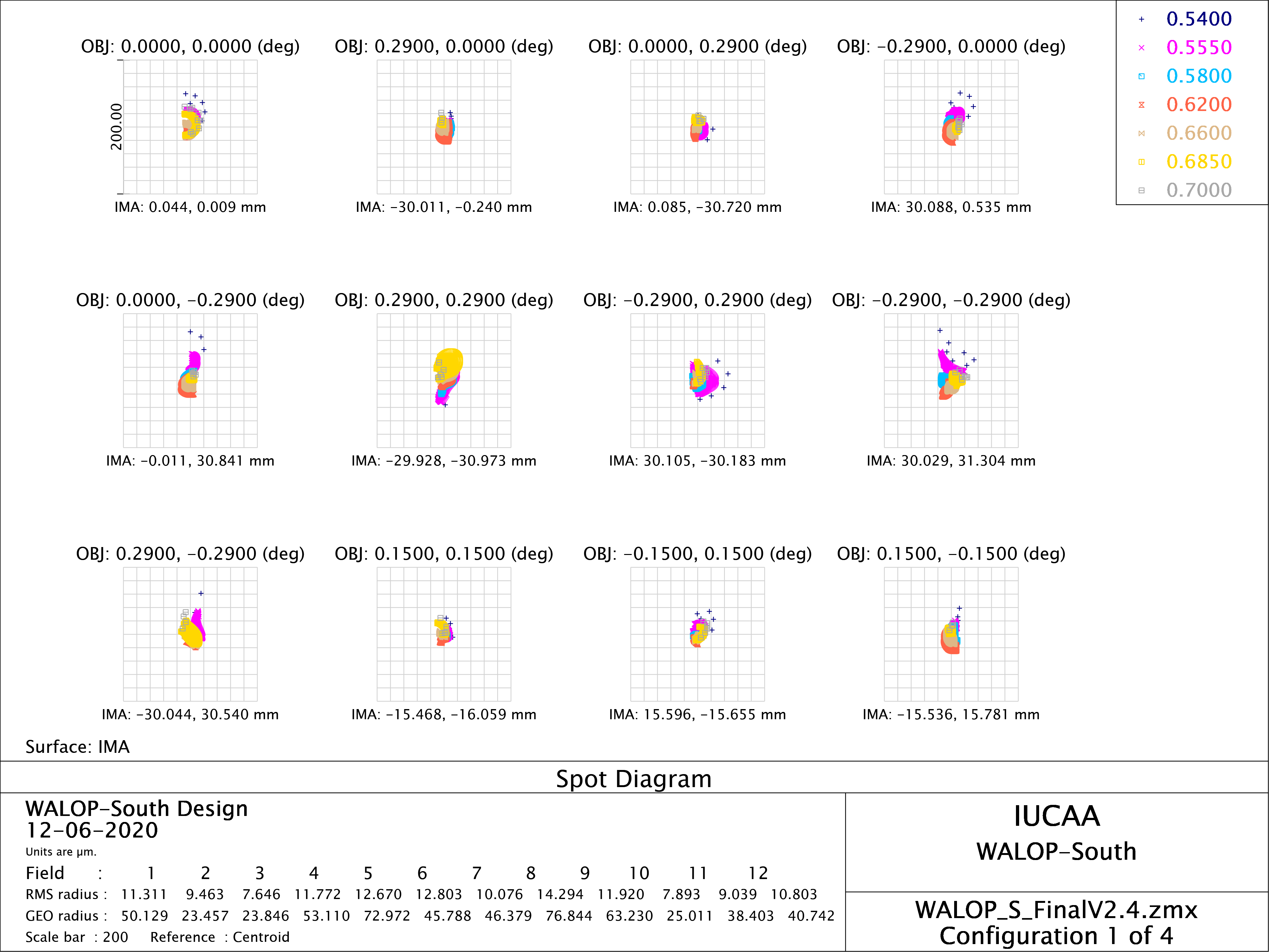}
    \caption{Spot diagram for one of the four cameras at the detector for different field points. Different colors represent different wavelengths as labeled in the image legend. RMS and GEO radius stand for the root-mean square and geometric radius of the spot diagrams, respectively. The optical performance of E1 and E2 beams are identical as they follow identical optical paths, and likewise for the O1 and O2 beams. Also, the O and E beams have similar spot diagram sizes(Table~\ref{MC_sim}). }
    \label{spot_diagram}
\end{figure}

Figure~\ref{spot_diagram} shows the spot diagram for one of the four cameras (O1 beam) at the detector for different field points. All the four beams have very similar spot diagrams. The O1 and O2 beams have identical optical paths and thus identical spot diagrams. Same is true for E1 and E2 beams. The averaged RMS (root mean squared) radii for the O and E beams ("Nominal spot radius" parameter in Table~\ref{MC_sim}) is 11.63 and 11.77~${\mu}m$ respectively. In comparison, the RMS radius for a 1.5 arcsecond FWHM Gaussian beam (median seeing at the Sutherland Observatory) at the detectors is 19.1~${\mu}m$.

A complete tolerance analysis of the optical system was done in Zemax using Monte Carlo (MC) simulations to estimate the expected deterioration in the spot sizes for the instrument and the required tolerances for the fabrication of optical and mechanical components of the system. Two compensators were defined- (a) separation between the primary and secondary telescope mirror, and (b) distance between the last camera lens and the detector of each camera. Table~\ref{MC_sim} shows the results of 20,000 MC simulations. The mean RMS spot radius for the O and E beams based on the simulations are 17.1 and 15.37~${\mu}m$ respectively for the O and E beams, which is smaller than the RMS radius for a 1.5 arcsecond FWHM Gaussian beam at the detectors (19.1~${\mu}m$). Thus we expect to obtain near seeing limited PSF at the detectors (for a comprehensive and quantified estimate of the instrument's expected imaging performance, refer to Paper I). Table~\ref{mech_toleraces} captures the required tolerances for the alignment of the optical assembly based on which the MC simulation results were obtained. The tolerance values are common for corresponding elements in all the four beams.

\begin{table}[ht!]
\centering
\begin{tabular}{|c|c|c|}
\hline
Parameter                & O-Beams             & E-Beams             \\ \hline
                         & RMS Spot Radius (${\mu}m$) & RMS Spot Radius (${\mu}m$) \\ \hline
Nominal Spot             & 11.63               & 11.77               \\ \hline
Root-Sum-Square          & 17.1                & 15.37               \\ \hline
MC Simulation Best Case  & 11.72               & 11.7                \\ \hline
MC Simulation Worst Case & 37.4                & 25.5                \\ \hline
MC Simulation Mean       & 17.54               & 15.72               \\ \hline
MC Simulation Std Dev    & 0.003               & 0.0018              \\ \hline
\end{tabular}
\caption{Results of Monte Carlo simulations based tolerance analysis for O and E beams. Root-Sum-Square radius is the RMS spot radius obtained if the offset in spot radius due to all mechanical and optical tolerances are added in quadrature.}
\label{MC_sim}
\end{table}

\begin{table}[htbp!]
\centering
\begin{tabular}{|c|c|c|c|}
\hline
\multirow{3}{*}{Lens   Name} & \multirow{3}{*}{Decentre~(${\mu}m$)} & \multirow{3}{*}{Axial~(${\mu}m$)} & \multirow{3}{*}{Tilt~(arcminute)} \\
                             &                                          &                                       &                                          \\
                             &                                          &                                       &                                          \\ \hline
Collimator Lens 1            & 50                                       & 200                                   & 3                                        \\ \hline
Collimator Lens 2            & 50                                       & 200                                   & 3                                        \\ \hline
Collimator Lens 3            & 50                                       & 200                                   & 3                                        \\ \hline
Collimator Lens 4            & 30                                       & 200                                   & 2                                        \\ \hline
Collimator Lens 5            & 50                                       & 200                                   & 1                                        \\ \hline
Collimator Lens 6            & 20                                       & 100                                   & 2                                        \\ \hline
Camera Lens 1                & 30                                       & 50                                    & 1                                        \\ \hline
Camera Lens 2                & 30                                       & 30                                    & 1                                        \\ \hline
Camera Lens 3                & 30                                       & 50                                    & 1                                        \\ \hline
Camera Lens 4                & 30                                       & 200                                   & 1                                        \\ \hline
Camera Lens 5                & 30                                       & 200                                   & 2                                        \\ \hline
Camera Lens 6                & 50                                       & 200                                   & 3                                        \\ \hline
Camera Lens 7                & 50                                       & 200                               & 3                                        \\ \hline
WPA                          & 50                                       & 100                                   & 5                                        \\ \hline
PBS                          & 50                                       & 100                                   & 5                                        \\ \hline
DC Prism 1                   & 50                                       & 100                                   & 5                                        \\ \hline
DC Prism 2                   & 50                                       & 100                                   & 5                                        \\ \hline
Fold Mirror                  & 50                                       & 100                                   & 5                                        \\ \hline
\end{tabular}

\caption{Tolerances on alignment of the optical elements of the WALOP-South instrument. The tolerance values are common for corresponding elements in all the four cameras.}
\label{mech_toleraces}
\end{table}

\section{Optomechanical Design}\label{optomech}

\subsection{Technical Requirements}
The requirements from the optomechanical design of the WALOP-South instrument are:

\begin{enumerate}
    \item Align and hold all the optical elements within the mechanical tolerances obtained from the tolerance analysis (Table~\ref{mech_toleraces}) of the instrument.
    \item Maintain alignment of the optics (within required tolerances) for the various possible different pointing orientations of the telescope, especially from zenith to up to $30^{\circ}$) from the horizon, i.e. airmass of 2 since most observations will be done in this telescope pointing window.
    \item Optics holders should exert minimal stresses on the glasses due to the mounting method as well as due to temperature changes at the telescope site during observations. Stress on glass leads to stress birefringence which will modify the polarization state of the light ray passing through the glass and will lead to instrumental polarization and cross-talk between the Stokes parameters.
    \item The maximum mass of instrument that can be mounted on the SAAO 1~m telescope is 150 Kgs; so the instrument mass should be under 150 Kg. 
    \item The instrument requires controlled motions for many subsystems. A list of all motion systems are captured in Table~\ref{control_goals} and elaborated in Section~\ref{mechanisms}. The optomechanical system should provide provisions for all such movements to the required accuracy.
    \item The instrument model should have provisions for mounting of all electrical connectors and control boxes such as CCD control boxes as well motion motion control boxes, taking into consideration locations where the connections are needed.
\end{enumerate}

\subsection{Overall Model}
\begin{figure}[ht!]
    \centering
    \fbox{\includegraphics[scale=0.5]{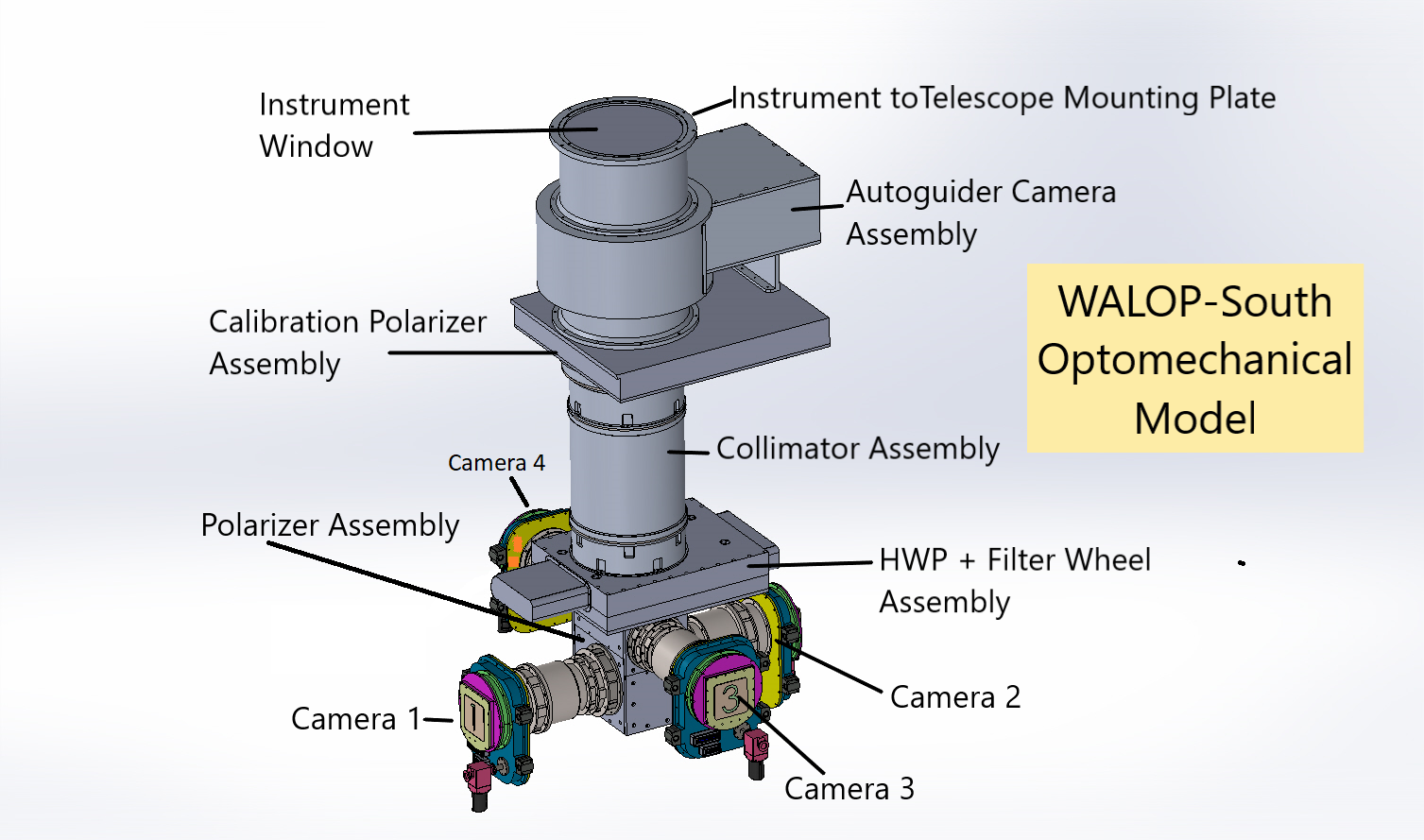}}
    \caption{The overall optomechanical model of the WALOP-South instrument, without electrical connectors and control boxes mounted. The various major subsystems in the model have been marked.}
    \label{optmech_shaded}
\end{figure}

Figure~\ref{optmech_shaded} shows the overall optomechanical model of the instrument, without electrical connectors and control boxes mounted. The different subsystems of the instrument are annotated in the image. 
The instrument begins with an instrument window to mechanically seal the instrument from the outside environment to prevent dust settling on optics surfaces that can lead to spurious polarization signals in polarimeters (refer to Figure 4 in RoboPol instrument paper\cite{robopol}). Before the main instrument, the auto-guider camera and calibration polarizer sub-assemblies are present. The optical design of the auto-guider camera described in Paper I. The calibration polarizer is a linear polarizer sheet which is provided for creating the polarization calibration model of the instrument. The main optomechanical model of the instrument, like the optical design, can be divided into following subsystems- (a) collimator barrel, (b) polarizer box and (c) four camera barrels. As the name suggests, the collimator and camera assemblies are in form of barrels (Section~\ref{barrel_design}) while the polarizer assembly consisting of the Wollaston Prism Assembly, wire-grid Polarization Beam-Splitters, Dispersion Corrector Prisms and the Fold Mirrors are all enclosed in a box from which the four camera barrels project in four directions. 
. 

\subsection{Barrel Design}\label{barrel_design}
\begin{figure}[htbp!]
    \centering
    \frame{\includegraphics[scale = 0.3]{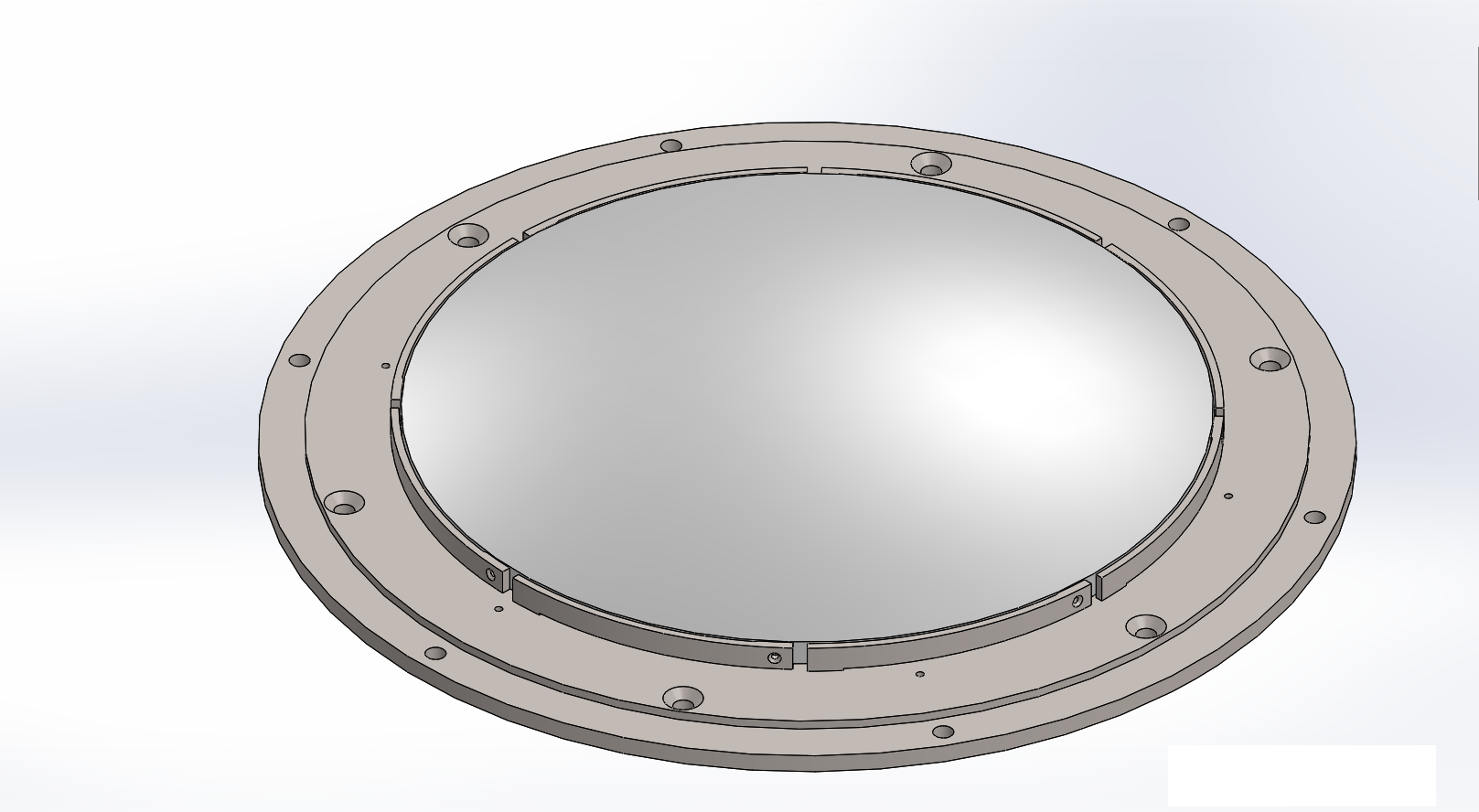}}
    \caption{A lens holder of the instrument with lens mounted on it. The lens is attached to the holder using the multiple flexures which are then glued to the lens around its rim (cylindrical face) at multiple places.}
    \label{lens_holder}
\end{figure}

\begin{figure}[htbp!]
    \centering
    \fbox{\includegraphics[scale = 0.5]{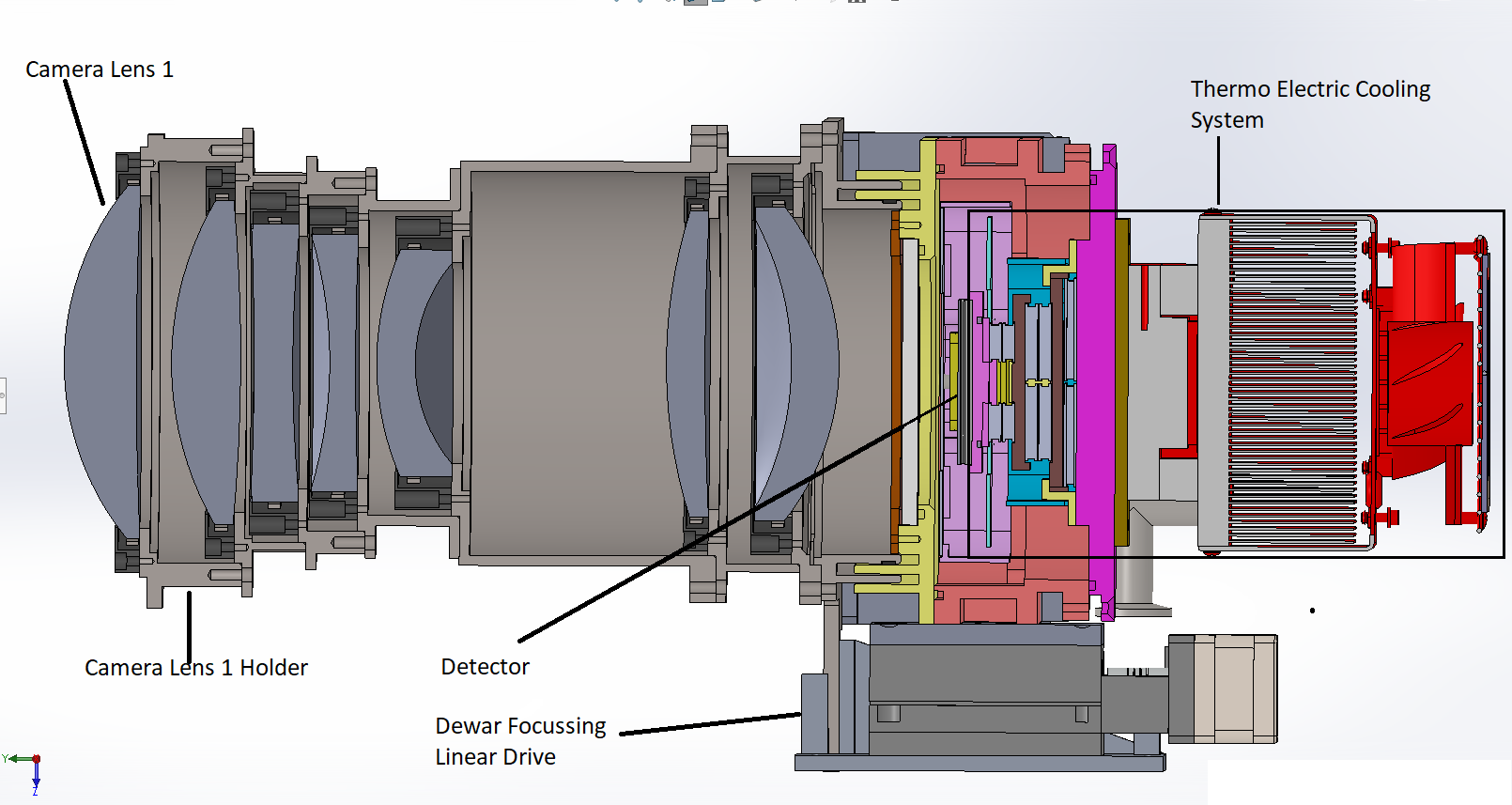}}
    \caption{Cross Section of one of the four camera barrels. Every lens is held in it's holder, which is then connected to the succeeding lens holder through a cylindrical spacer. For additional protection, each lens has retainer made from a soft material (Teflon) attached to it. In this barrel, to obtain better alignment accuracy, the lens holder and spacer for all the individual lenses have been integrated into one component. }
    \label{camera_barrel_cross_section}
\end{figure}

The lenses are held in their individual holders using flexure based lens mounts which are the widely used to achieve and maintain high accuracy alignments\cite{Vukobratovich_flexure}. Figure~\ref{lens_holder} shows a lens holder with a lens mounted on it. The lens is attached to the holder using the multiple flexures which are glued to the lens around its rim (cylindrical face). The collimator and the camera barrels are made by placing the lens holders in sequence with cylindrical spacers. Figure~\ref{camera_barrel_cross_section} show the images of the one of the four camera barrels of WALOP-South instrument. Following design decisions were made in the barrel design:

\begin{enumerate}
    \item Most lens mounts, except for the largest lenses (collimator lens 1 and 2), the lens mount material is made of Aluminium-6061 alloy; for the larger lenses, Titanium 6Al-4V alloy is used. The CTE (coefficient of thermal expansion) of Titanium 6Al-4V alloy is $8.9\times10^{-6}$, which is close to the CTE of most glasses ($7-10\times10^{-6}$), while the CTE of Aluminium-6061 alloy is $24\times10^{-6}$ and is farther away from that of most lens materials. While Titanium 6Al-4V is apt for reducing mechanical stresses on the glasses due to temperature changes, it is a heavier material $4.4~{g/cc}$ than Aluminium-6061 alloy ($2.7~{g/cc}$). Additionally, Aluminium-6061 is cheaper and easier to procure, and more importantly easy to machine to the tight tolerances required by us. So barring the first two lenses where we expect large thermal stresses to arise due to the larger aperture of the lenses, all other lens mounts have been made of Aluminium-6061 alloy. 
    \item In optical lenses requiring high accuracy alignment, the spacer and lens mount have been combined so to be made a single mechanical component, reducing additional mechanical misalignment.
\end{enumerate}

\subsection{Motion Mechanisms and Control Systems}\label{mechanisms}
Table~\ref{control_goals} lists all the control systems in the WALOP-South instrument. The calibration linear polarizer sheet at the beginning of the instrument and the calibration Half-Wave Plate at the pupil need to have rotation motions when in the optical path with the provision of being moved out of the optical path when not in use. The filter wheel, placed at the pupil, has 4 filters mounted on a linear stage, and any of the four can be placed in the optical path by the linear motion of the stage. The guider camera (refer to Paper I for guider camera optical design) has two linear stages on which the entire camera optics is mounted. The x-y motion of these two stages is used to patrol an effective field of view of 540 square arcminutes. There is a provision for placing a filter in the auto-guider camera's optical path by in-out motion. A common shutter for all the four cameras is placed inside the collimator barrel which is controlled electronically to open and close. 

\par The Wollaston Prism Assembly is the most delicate and temperature sensitive optical component in the instrument (refer to Paper I for details). While it has been cemented with the flexible Norland-65 cement, chosen such that it can withstand all temperature conditions at SAAO without mechanical fracture, we will temperature control the Wollaston Prism Assembly at $23^{\circ}~C$, the temperature at which the assembly has been cemented. This provides an additional measure for safety against thermal stresses in the Wollaston Prism Assembly to prevent damage as well as reduce stress birefringence which can affect it's polarimetric performance.

\begin{table}[htbp!]
\centering
\begin{tabular}{|c|c|c|c|c|c|c|}
\hline
Serial No. & Subsystem                                                                   & \begin{tabular}[c]{@{}c@{}}Control\\  Required\\ 1\end{tabular}   & \begin{tabular}[c]{@{}c@{}}Control\\  Required\\ 2\end{tabular} & \begin{tabular}[c]{@{}c@{}}Control \\ Required\\ 3\end{tabular} & \begin{tabular}[c]{@{}c@{}}No. of \\ Motions\end{tabular} & Location                                                               \\ \hline
1          & \begin{tabular}[c]{@{}c@{}}Calibration \\ Polarizer\end{tabular}            & In-out                                                            & Rotation                                                        & -                                                               & 2                                                         & \begin{tabular}[c]{@{}c@{}}Guider + Cal.\\ Polarizer Box\end{tabular}  \\ \hline
2          & \begin{tabular}[c]{@{}c@{}}Auto-Guider \\ Camera \\ Patrolling\end{tabular} & \begin{tabular}[c]{@{}c@{}}X-direction \\ motion\end{tabular}     & \begin{tabular}[c]{@{}c@{}}Y-direction\\  motion\end{tabular}   & -                                                               & 2                                                         & \begin{tabular}[c]{@{}c@{}}Guider + Cal.\\ Polarizer Box\end{tabular}  \\ \hline
3          & \begin{tabular}[c]{@{}c@{}}Auto-Guider\\  Camera\end{tabular}               & Filter in-out                                                     & \begin{tabular}[c]{@{}c@{}}Exposure \\ Control\end{tabular}     & -                                                               & 1                                                         & \begin{tabular}[c]{@{}c@{}}Guider + Cal. \\ Polarizer Box\end{tabular} \\ \hline
4          & \begin{tabular}[c]{@{}c@{}}Half Wave\\  Plate\end{tabular}                  & In-out                                                            & Rotation                                                        & -                                                               & 2                                                         & \begin{tabular}[c]{@{}c@{}}HWP + Filter \\ Wheel Box\end{tabular}      \\ \hline
5          & Filter Wheel                                                                & Rotation                                                          & -                                                               & -                                                               & 1                                                         & \begin{tabular}[c]{@{}c@{}}HWP + Filter \\ Wheel Box\end{tabular}      \\ \hline
6          & Shutter                                                                     & Open/Close                                                        & -                                                               & -                                                               & -                                                         & \begin{tabular}[c]{@{}c@{}}Collimator \\ Barrel\end{tabular}           \\ \hline
7          & \begin{tabular}[c]{@{}c@{}}Wollaston \\ Prism \\ Assembly\end{tabular}      & \begin{tabular}[c]{@{}c@{}}Temperature \\ Control\end{tabular}    & -                                                               & -                                                               & -                                                         & Polarizer Box                                                          \\ \hline
8          & Dewar 1                                                                     & \begin{tabular}[c]{@{}c@{}}Linear Focus \\ Mechanism\end{tabular} & \begin{tabular}[c]{@{}c@{}}Temperature \\ Control\end{tabular}  & \begin{tabular}[c]{@{}c@{}}CCD \\ Readout\end{tabular}          & 1                                                         & \begin{tabular}[c]{@{}c@{}}Camera 1 \\ Dewar\end{tabular}              \\ \hline
9          & Dewar 2                                                                     & \begin{tabular}[c]{@{}c@{}}Linear Focus \\ Mechanism\end{tabular} & \begin{tabular}[c]{@{}c@{}}Temperature \\ Control\end{tabular}  & \begin{tabular}[c]{@{}c@{}}CCD \\ Readout\end{tabular}          & 1                                                         & \begin{tabular}[c]{@{}c@{}}Camera 2 \\ Dewar\end{tabular}              \\ \hline
10         & Dewar 3                                                                     & \begin{tabular}[c]{@{}c@{}}Linear Focus \\ Mechanism\end{tabular} & \begin{tabular}[c]{@{}c@{}}Temperature \\ Control\end{tabular}  & \begin{tabular}[c]{@{}c@{}}CCD \\ Readout\end{tabular}          & 1                                                         & \begin{tabular}[c]{@{}c@{}}Camera 3 \\ Dewar\end{tabular}              \\ \hline
11         & Dewar 4                                                                     & \begin{tabular}[c]{@{}c@{}}Linear Focus \\ Mechanism\end{tabular} & \begin{tabular}[c]{@{}c@{}}Temperature\\  Control\end{tabular}  & \begin{tabular}[c]{@{}c@{}}CCD \\ Readout\end{tabular}          & 1                                                         & \begin{tabular}[c]{@{}c@{}}Camera 4 \\ Dewar\end{tabular}              \\ \hline
\end{tabular}
\caption{Details of the various control systems used in WALOP-South instrument.}
\label{control_goals}
\end{table}

Each dewar houses a $4k\times4k$ E2V CCD which is maintained at $-100~C$ through thermo-electric cooling system. The CCDs are read-out using controllers developed in-house in the IUCAA lab. Each dewar has linear motion drive to allow motion along the optical axis for focusing individual cameras (used as compensator in tolerance analysis).

\section{Current Status}
We are finalizing the optomechanical design of the instrument after which we will proceed towards assembly and testing of the instrument in the lab. The instrument is scheduled for commissioning in the year 2021.

\acknowledgments 
The PASIPHAE program is supported by grants from the European Research Council (ERC) under grant agreement No 771282 and No 772253, from the National Science Foundation, under grant number AST-1611547 and the National Research Foundation of South Africa under the National Equipment Programme. This project is also funded by an infrastructure development grant from the Stavros Niarchos Foundation and from the Infosys Foundation.

\bibliography{SPIE2020_Proceedings} 

\begin{thebibliography}{1}

\bibitem{tassis2018pasiphae}
Tassis, K., Ramaprakash, A.~N., Readhead, A. C.~S., Potter, S.~B., Wehus,
  I.~K., Panopoulou, G.~V., Blinov, D., Eriksen, H.~K., Hensley, B., Karakci,
  A., Kypriotakis, J.~A., Maharana, S., Ntormousi, E., Pavlidou, V., Pearson,
  T.~J., and Skalidis, R., ``Pasiphae: A high-galactic-latitude, high-accuracy
  optopolarimetric survey,'' (2018).

\bibitem{Heiles_2000}
Heiles, C., ``9286 {STARS}: {AN} {AGGLOMERATION} {OF} {STELLAR} {POLARIZATION}
  {CATALOGS},'' {\em The Astronomical Journal}~{\bf 119},  923--927 (feb 2000).

\bibitem{Panopoulou_2019}
Panopoulou, G.~V., Tassis, K., Skalidis, R., Blinov, D., Liodakis, I.,
  Pavlidou, V., Potter, S.~B., Ramaprakash, A.~N., Readhead, A. C.~S., and
  Wehus, I.~K., ``Demonstration of magnetic field tomography with starlight
  polarization toward a diffuse sightline of the ism,'' {\em The Astrophysical
  Journal}~{\bf 872},  56 (Feb 2019).

\bibitem{robopol}
Ramaprakash, A.~N., Rajarshi, C.~V., Das, H.~K., Khodade, P., Modi, D.,
  Panopoulou, G., Maharana, S., Blinov, D., Angelakis, E., Casadio, C.,
  Fuhrmann, L., Hovatta, T., Kiehlmann, S., King, O.~G., Kylafis, N.,
  Kougentakis, A., Kus, A., Mahabal, A., Marecki, A., Myserlis, I., Paterakis,
  G., Paleologou, E., Liodakis, I., Papadakis, I., Papamastorakis, I.,
  Pavlidou, V., Pazderski, E., Pearson, T.~J., Readhead, A. C.~S., Reig, P.,
  Słowikowska, A., Tassis, K., and Zensus, J.~A., ``{RoboPol: a four-channel
  optical imaging polarimeter},'' {\em Monthly Notices of the Royal
  Astronomical Society}~{\bf 485},  2355--2366 (02 2019).

\bibitem{HOWPol}
Kawabata, K.~S., Nagae, O., Chiyonobu, S., Tanaka, H., Nakaya, H., Suzuki, M.,
  Kamata, Y., Miyazaki, S., Hiragi, K., Miyamoto, H., Yamanaka, M., Arai, A.,
  Yamashita, T., Uemura, M., Ohsugi, T., Isogai, M., Ishitobi, Y., and Sato,
  S., ``{Wide-field one-shot optical polarimeter: HOWPol},'' in [{\em
  Ground-based and Airborne Instrumentation for Astronomy
  II}{\nolinebreak\hspace{0.1em}]},  McLean, I.~S. and Casali, M.~M., eds.,
  {\bf 7014},  1585 -- 1594, International Society for Optics and Photonics,
  SPIE (2008).

\bibitem{salt_commisioning}
Potter, S.~B., Nordsieck, K., Romero-Colmenero, E., Crawford, S., Vaisanen, P.,
  Éric Depagne, Buckley, D., Koeslag, A., Brink, J., Hetlage, C., Browne, K.,
  Crause, L., Schier, A., and Allington, J., ``{Commissioning the polarimetric
  modes of the Robert Stobie spectrograph on the Southern African Large
  Telescope},'' in [{\em Ground-based and Airborne Instrumentation for
  Astronomy VI}{\nolinebreak\hspace{0.1em}]},  Evans, C.~J., Simard, L., and
  Takami, H., eds.,  {\bf 9908},  810 -- 817, International Society for Optics
  and Photonics, SPIE (2016).

\bibitem{Vukobratovich_flexure}
Vukobratovich, D. and Richard, R.~M., ``{Flexure Mounts For High-Resolution
  Optical Elements},'' in [{\em Optomechanical and Electro-Optical Design of
  Industrial Systems}{\nolinebreak\hspace{0.1em}]},  Bieringer, R.~J. and
  Harding, K.~G., eds.,  {\bf 0959},  18 -- 36, International Society for
  Optics and Photonics, SPIE (1988).

\end{thebibliography}
\bibliographystyle{spiebib} 

\end{document}